# Progression models for repeated measures: Estimating novel treatment effects in progressive diseases

Lars Lau Raket, PhD[1, 2]

[1]Novo Nordisk A/S, Vandtårnsvej 108, 2860 Søborg, Denmark
[2]Clinical Memory Research Unit, Department of Clinical Sciences, Lund University, Lund, Sweden

**Abstract**

Mixed Models for Repeated Measures (MMRMs) are ubiquitous when analyzing outcomes of clinical trials. However, the linearity of the fixed-effect structure in these models largely restrict their use to estimating treatment effects that are defined as linear combinations of effects on the outcome scale. In some situations, alternative quantifications of treatment effects may be more appropriate. In progressive diseases, for example, one may want to estimate if a drug has cumulative effects resulting in increasing efficacy over time or whether it slows the time progression of disease. This paper introduces a class of nonlinear mixed-effects models called Progression Models for Repeated Measures (PMRMs) that, based on a continuous-time extension of the categorical-time parametrization of MMRMs, enables estimation of novel types of treatment effects, including measures of slowing or delay of the time progression of disease. Compared to conventional estimates of treatment effects where the unit matches that of the outcome scale (e.g. 2 points benefit on a cognitive scale), the time-based treatment effects can offer better interpretability and clinical meaningfulness (e.g. 6 months delay in progression of cognitive decline). The PMRM class includes conventionally used MMRMs and related models for longitudinal data analysis, as well as variants of previously proposed disease progression models as special cases. The potential of the PMRM framework is illustrated using both simulated and historical data from clinical trials in Alzheimer's disease with different types of artificially simulated treatment effects. Compared to conventional models it is shown that PMRMs can offer substantially increased power to detect disease-modifying treatment effects where the benefit is increasing with treatment duration.

# Introduction

Treatment effects can take many forms. In progressive or degenerative diseases, where patients are expected to deteriorate over time, different types of interventions may have very different effects. For example, some interventions reduce certain symptoms of disease without changing the overall course; some interventions delay progression by intervening in the biological processes that cause disease; and some interventions may even be curative or restore lost function.

The mixed model for repeated measures (MMRM)[1, 2] offers a very general approach for analyzing data from clinical trials with minimal assumptions on the effect of treatment on outcome measures. The defining characteristic of the MMRM is the categorical modeling of time through a visit variable which allows approximation of any shape of trajectory of the outcome measure.

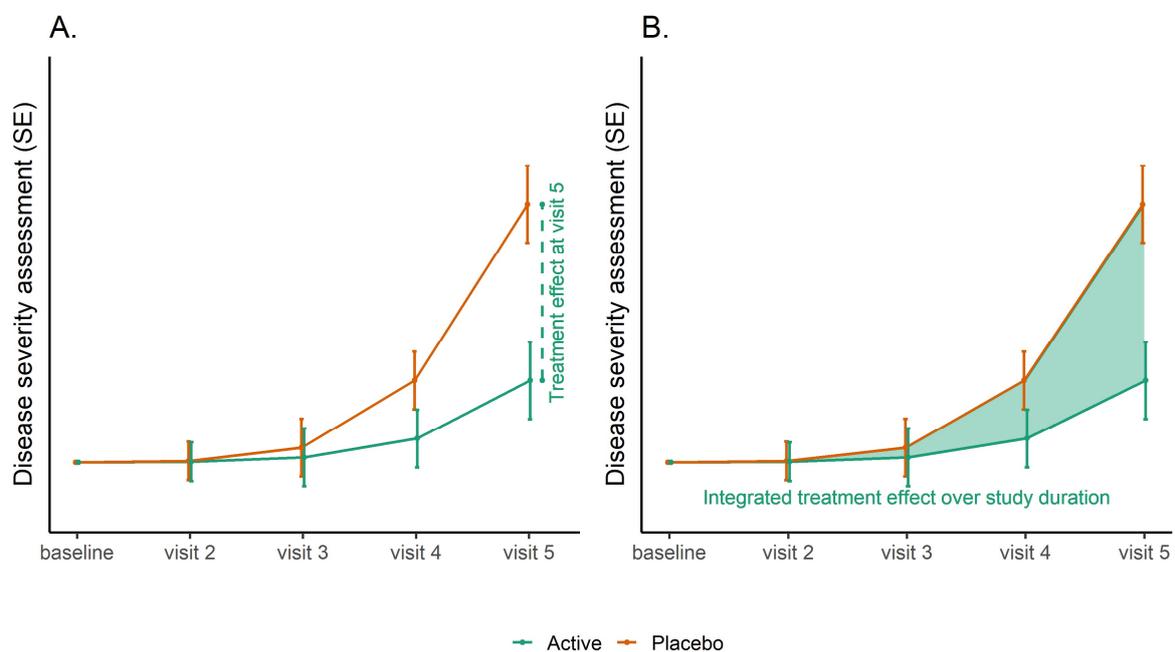

**Figure 1.** Trajectories of active and placebo groups estimated using an MMRM in a hypothetical study. A. The treatment effect is quantified as the estimated difference between active and placebo at visit 5. B. The treatment effect is quantified as the estimated difference in area (shaded) under the curves of active and placebo treatment over the study duration. *SE*: standard error

Consider the hypothetical trial outcomes in Figure 1. The MMRM can address questions such as "is there a treatment effect at the final visit and what is its magnitude on the outcome scale?" by testing and evaluating contrasts of parameters (Figure 1A). The MMRM can also assess other types of treatment effects that can be expressed as linear combinations of parameters. For example, the cumulative benefit of a treatment over the course of a study may be evaluated by the differences in area under the curves spanned by the mean parameters which can be approximated using the trapezoidal rule on estimated model parameters and between-visit durations[3] (Figure 1B). However, such estimates of treatment effect may not always offer the most meaningful or efficient quantification of the impact of an intervention.

*When are conventional analysis methods insufficient?*

One may wish to quantify treatment effects differently when treatment effects differences are hard to interpret or if treatment is expected to alter the time course of disease.

Consider a slowly progressing diseases such as Alzheimer's disease or Parkinson's disease. A treatment that slows the progression of early-stage disease is likely to only show a modest benefit during a clinical trial of typical duration (12-24 months) because very limited decline is expected in the reference group.[4] If a treatment slows the time-progression of disease by 25%, it will have delayed disease progression by 6 months after 24-months of treatment, but if the average decline in study outcome measure (e.g. cognitive or motor function) is only very limited during this 24-month period, a conventional MMRM analysis will find that the treatment effect measured as points on the outcome measure is very small, and the clinical relevance may be deemed questionable. On the other hand, a 6-month delay of disease progression in the early stages of these diseases would likely be considered both meaningful and interpretable by prescribers, patients, and care partners. As this example illustrates, there are situations where alternative ways to show and quantify benefits over time may be more meaningful than treatment differences on the outcome scale. As an added benefit, more appropriate modeling of such treatment effects could offer greater power to detect treatment effects if the models make better use information from the full course of the trial in the estimation of the treatment effect.

*Previous use of progression models in clinical trials of progressive diseases*

Novel disease progression models have emerged as a promising approach for detecting and quantifying treatment effects in several recent trials in Alzheimer's disease. For example, disease progression models have been used to analyze the outcomes in the DIAN-TU study in dominantly inherited Alzheimer's disease,[5,6] the TRAILBLAZER-ALZ study in early Alzheimer's disease,[7] and a similar modeling approach was developed for the EPAD study.[8,9] The mentioned models are all Bayesian, rely on an assumption of proportionality of decline in active and placebo arms, have subject-level random intercepts and assume a simple error structures with independent identically normally-distributed errors. The models differ slightly in choices of priors and in terms of how they model disease progression relative to the baseline visit due to the differences in study designs and populations.

While simulation studies have showed that such modeling approaches can offer substantial power gains compared to traditional MMRM analyses,[5] the assumptions of the models in combination with the simplicity of the error structure may impact model results or lead to convergence problems when used on real data that do not fit these assumptions as seen in the DIAN-TU study.[6] Furthermore, the treatment effects should be interpreted as percent reduction in decline on an outcome measure over the course of the study, but observed reduction in decline depends both on the sensitivity of the outcome measure, patient population and trial duration. Therefore, reduction in decline tends to be an unstable measure over time, and even if a treatment slows the progression of disease at a consistent rate, it is unlikely that different outcome measures will show consistent reductions in decline, thus limiting the interpretability of such estimates.

*Contribution of the present paper*

This paper presents a new class of models, called Progression Models for Repeated Measures (PMRMs). PMRMs provide a framework for efficient estimation of interpretable treatment effects from clinical trials where the aim is to slow or halt disease progression. This class of models combines the flexibility and robustness of the MMRM with the potential of low-dimensional parametrizations of treatment effects and the ability of specifying treatment effects that alter the time course of disease.

These types of models are particularly relevant in proteinopathies such as Alzheimer's and Parkinson's disease, where potential disease-modifying therapies that target the defining pathological hallmarks of disease are often expected to slow the progression of symptomatic disease.[10] In such cases, one would expect treatment to result in similar slowing of progression across endpoints that represents different aspects of disease symptoms. Due to the common unit of time of treatment effects in the time-based PMRMs, they naturally allow comparison of results across outcomes or even estimation of a simultaneous treatment effects based on multiple outcomes.

In the following, we first introduce the types of treatment effects being studies. Then the PMRM modeling framework is introduced and illustrated using both simulated and historical trial data from Alzheimer's disease studies. Through these examples, it is demonstrated that in addition to offering more interpretable quantifications of treatment effects, PMRMs can offer substantial power gains in some situations.

## Symptomatic and disease-modifying treatment effects

While there are endless ways in which treatment effects can manifest on outcome measures, one can generally distinguish between treatment effects that are symptomatic versus disease modifying.[11] Slowing, stopping, or reversing the course of progressive disease would conventionally be thought of as disease-modifying treatment effects, as opposed to symptomatic treatment effects. Symptomatic treatments improve symptom presentation but not the underlying cause, whereas disease-modifying treatments affect the underlying cause which in turn impact the symptomatic manifestation that is quantified in study outcome measures.

In the following, we will consider the six stylized examples of treatment effects on continuous outcome measures illustrated in Figure 2. The effects are: symptomatic improvements that builds over an initial period and after that manifests as a constant improvement measured on the outcome scale (Figure 2A) or a similar improvement that fades with time (Figure 2B); a proportional reduction of decline on the outcome scale (Figure 2C); a time-delay of the progression (Figure 2D); a proportional slowing of the time-progression (Figure 2E); and a non-proportional slowing of the time-progression (Figure 2F). It is worth noting that the three last types of treatment effects may show decreasing efficacy along the course of treatment as measured on the (vertical) outcome scale, depending on the shape of the natural trajectory of disease. This, however, does not necessarily correspond to a reduced benefit – if the drug proportionally slows the (horizontal) progression of disease, patients will gain the cumulative benefit of more time in milder disease states throughout the study, but if the outcome measure reaches a ceiling or otherwise loses its sensitivity to change as patients progress during the study, this may be observed as a decrease in vertical distance between the curves, even if the horizontal distance keeps increasing.

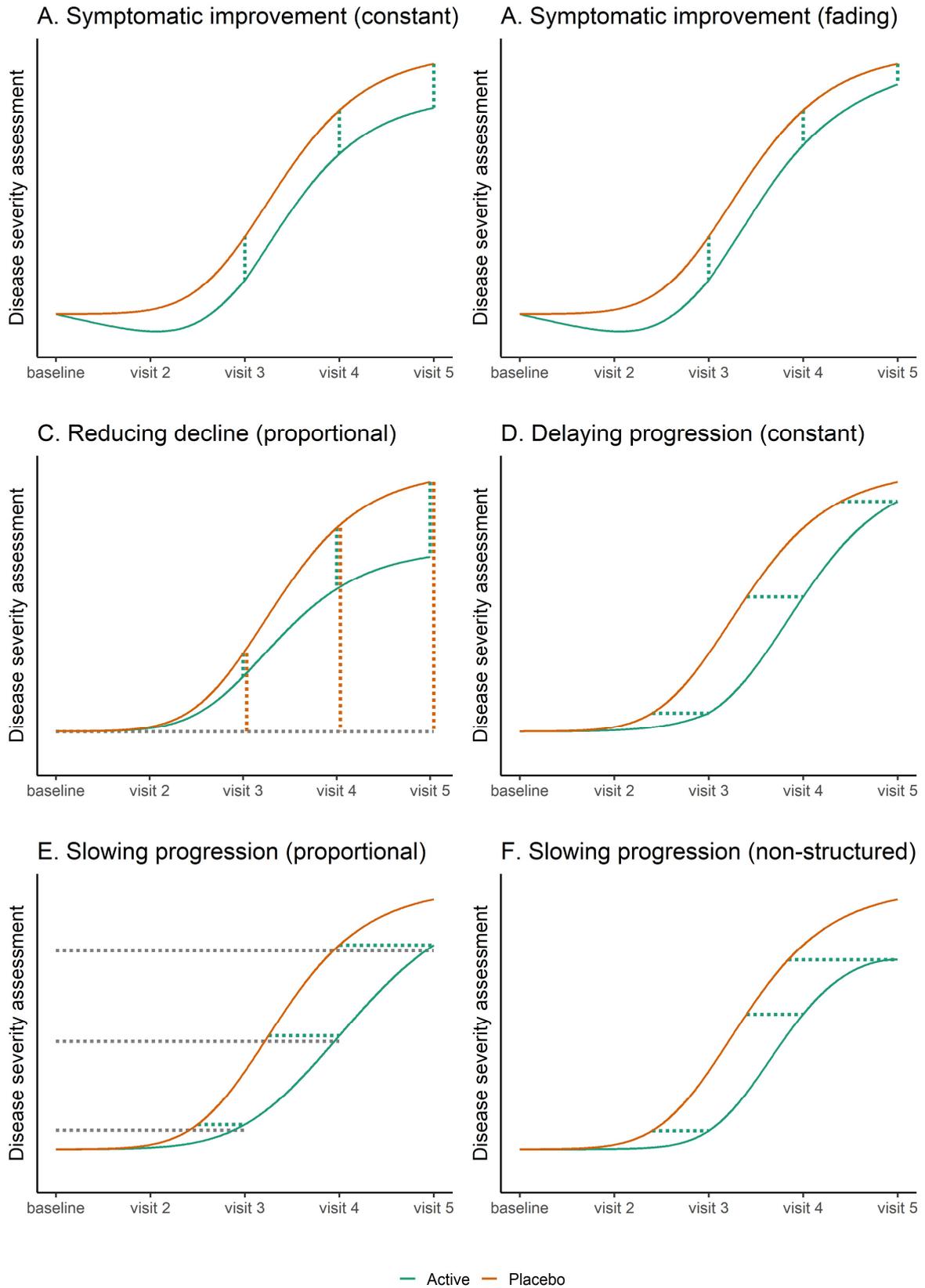

**Figure 2.** Six examples of treatment effects. Dotted green lines show the treatment differences used to generate the curves. A. The symptomatic treatment effect builds from baseline to visit 3 and manifests as a constant vertical difference between the curves from visit 3 to 5. B. The symptomatic treatment effect builds from baseline to visit 3 and then decreases slightly until visit 5. C. The treatment effect

manifests as a proportional reduction of the (vertical) decline observed in the placebo group, the ratio of the treatment differences (dotted green lines) and the placebo decline (dotted red lines) is constant across at all visits. D. The delay in progression builds from baseline to visit 3 and manifests as a constant (horizontal) delay from visit 3 to 5. E. The slowing of progression manifests as proportionally slower disease progression in the active arm where the ratio of the treatment-related delay in progression at each visit (dotted green lines) and the treatment duration (dotted grey lines) is constant. F. The slowing of progression manifests with similar delays at visits 3 and 4 for the active arm which then increases substantially up to visit 5.

## Mixed models for repeated measures, longitudinal data analysis models and a continuous-time extension

Let $y_{ij}$ denote the outcome measure of subject $i$ at visit $j$, $i = 1, \ldots, n$, $j = 0, 1, \ldots, m$, and let $\boldsymbol{y}_i$ denote the $(m + 1)$-dimensional vector of outcomes for subject $i$. The general linear mixed model can be written as

$$\boldsymbol{y}_i = \boldsymbol{X}_i \boldsymbol{\alpha} + \boldsymbol{Z}_i \boldsymbol{b}_i + \boldsymbol{e}_i \tag{1}$$

where $\boldsymbol{X}_i \in R^{(m+1) \times q}$ denotes the fixed-effect design matrix for subject $i$ with associated coefficients $\boldsymbol{\alpha} \in R^q$. $\boldsymbol{Z}_i \in R^{(m+1) \times p}$ denotes the random-effect design matrix with associated random coefficients $\boldsymbol{b}_i \sim N_p(0, \boldsymbol{D})$ and $\boldsymbol{e}_i$ is a vector of zero-mean error terms $\boldsymbol{e}_i \sim N_{m+1}(0, \boldsymbol{R})$.

In the following, we will model data using a constrained longitudinal data analysis (cLDA) approach where baseline measures are considered dependent variables in the model with assumed identical mean values across treatment arms due to randomization.[12] This approach differs slightly from the conventional use of MMRMs where change from baseline is modeled and the analysis is adjusted for baseline values through a visit-by-baseline value interaction covariate. When no data is missing, point estimates of treatment differences are identical with the two approaches, but when data is missing the cLDA approach may have some slight advantages.[13]

For simplicity, assume that we are modeling data from a randomized trial with a single active arm and a placebo arm. To keep the notation simple, we consider the following simplified version of model (1), whose fixed-effect design only describes the means of treatment arms over visits, without any explicit random effects

$$\boldsymbol{y}_i = \boldsymbol{X}_i \boldsymbol{\alpha} + \boldsymbol{e}_i \tag{2}$$

where $\boldsymbol{\alpha} = (\alpha_0, \alpha_{1,\text{placebo}}, \ldots, \alpha_{m,\text{placebo}}, \alpha_{1,\text{active}}, \ldots, \alpha_{m,\text{active}})^\top \in R^{2m+1}$ is the vector of mean parameters at baseline and the follow-up visits for the placebo and active arms and $\boldsymbol{X}_i$ is the subject-level design matrix that maps the mean parameters to the observations corresponding to the correct visits and treatment allocation. A visit-level formulation of this model is

$$y_{ij} = \alpha_{j,\text{treatment}(i)} + e_{ij}$$

where $\alpha_{j,\text{treatment}(i)}$ describes the mean outcome at visit $j$ for the treatment group that subject $i$ is randomized to, with the baseline constraint $\alpha_0 = \alpha_{0,\text{active}} = \alpha_{0,\text{placebo}}$.

## From categorical to continuous time

Treating visits as categorical variables makes the MMRM and cLDA model robust to all types of trajectories, but also makes it harder to model relations between treatment and time or effects of visit timing. While these models makes no assumption to what happens between visits, one would conventionally display estimates graphically by connecting the estimated mean values at the visits by lines, and one would think of these trajectories as representing a reasonable approximation of what goes on between visits. As we will see in the following, this functional representation can be considered the (restricted) maximum-likelihood estimate of a model with additional assumptions. Let $t_{ij}$ denote a continuous time variable describing the time since baseline of subject $i$'s visit $j$ with baseline time $t_{i0} = 0$. One can make a general nonlinear extension of model (2) of the form

$$y_{ij} = f(t_{ij}; \boldsymbol{X_i \alpha}) + e_{ij} \qquad (3)$$

where $f$ is a function that maps time to mean outcome value. The shape of $f$ is determined by the parameter vector $\boldsymbol{X_i \alpha}$.

The cLDA model (2) can be represented as such a continuous-time model: Choosing $t_{ij} = t_j$ to represent the scheduled visit times and modeling $f$ in (2) as an interpolating function of $\boldsymbol{X_i \alpha}$ with anchor points at the scheduled visits $t_0, t_1, \ldots, t_m$ will lead to exactly the same likelihood function and inference as the original model, however, the model estimates will be the full trajectories of each group. Figure 3 shows hypothetical cLDA point estimates along with the trajectories estimated by modeling $f$ as a zero-order (piecewise constant) spline, as a linear spline, as a natural cubic . From a likelihood inference perspective, these models are all equivalent to the cLDA model since the model estimates are only different in-between visits where no data is available. Thus, the choice of interpolation function should be based on prior knowledge of how one would expect the outcomes to behave between visits. In practice, one would often expect mean trajectories of outcome measures to be continuous and smooth, in which case a natural cubic spline would offer a reasonable choice of interpolation function.

The continuous-time extension also offers the opportunity to use actual visit times instead of the scheduled times in which case the likelihood function would differ from the cLDA model (2).

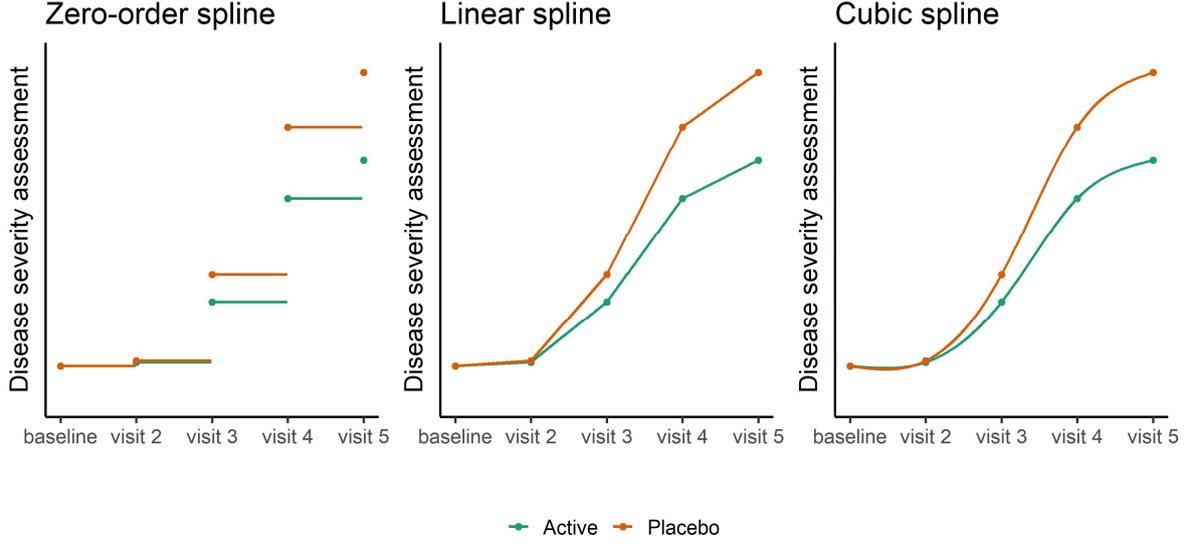

**Figure 3.** Three different examples of continuous-time modeling of hypothetical data collected over fixed visits. For the zero-order spline interpolation, the active and placebo trajectories are exactly on top of each other until visit 2 due to the assumption of a common baseline mean value.

## Progression Models for Repeated Measures

A PMRM is a model for which data in the placebo group is modeled as a continuous-time extension model (3) with the shape of the mean trajectory $f_0$ being determined by the parameters $\boldsymbol{\alpha} = (\alpha_0, \alpha_{1,\text{placebo}}, \dots, \alpha_{m,\text{placebo}})^\top$. The relation between the shape of the mean trajectory in the placebo group $f_0$ and the mean values of the active groups is described by a function $g$ and the parameter vector $\boldsymbol{\beta} = (\beta_0, \beta_1, \dots, \beta_k)^\top$

$$y_{ij} = g(t_{ij}; \boldsymbol{\alpha}, \boldsymbol{V}_{ij}\boldsymbol{\beta}) + e_{ij} \tag{4}$$

where $\boldsymbol{V}_{ij} \in R^{r \times k+1}$ is the design matrix that appropriately maps the $\beta$-parameters to an $r$-dimensional vector used to model the mean for the corresponding subject and visit. In the following sections, we will illustrate this concept through different examples of PMRMs.

### Modeling proportional reduction in decline

While model (2) can fully capture a treatment effect that manifests as a proportional reduction in decline (Figure 2C), it does so with redundant parameters since the proportionality assumption translates into the following constraints on the parameters

$$\alpha_{j,\text{active}} - \alpha_0 = \beta \cdot (\alpha_{j,\text{placebo}} - \alpha_0), \qquad \text{for all } j = 1, \dots, m.$$

For a study with $m = 6$ post baseline visits, the cLDA model (2) uses 13 free parameters to describe the mean values for the two groups in the study, whereas the constrained model only needs 8 free parameters with only a single of these describing the treatment effect.

The above constraint can be modeled with a PMRM (4) by choosing

$$g(t_{ij}; \boldsymbol{\alpha}, \boldsymbol{V}_{ij}\boldsymbol{\beta}) = \boldsymbol{V}_{ij}\boldsymbol{\beta} \cdot (f_0(t_{ij}; \boldsymbol{\alpha}) - \alpha_0) + \alpha_0 \tag{5}$$

where $f_0$ is the mean trajectory of the placebo group, $\boldsymbol{\beta} = (1, \beta)^\top$ and

$$\boldsymbol{V}_{ij} = \begin{cases} (1,0) & \text{for treatment}(i) = \text{placebo} \\ (0,1) & \text{for treatment}(i) = \text{active.} \end{cases}$$

Compared to previously used Bayesian disease progression models[5-7], the above PMRM is frequentist and has the additional flexibility of allowing an unstructured covariance of error terms. A similar frequentist extension of previous Bayesian disease progression models that leads to an identical model was recently proposed by Wang and colleagues.[14]

## Modeling time-based changes in disease progression

As opposed to *decline* which describes the worsening observed in an outcome measure over time, *progression* refers to the evolution of disease that causes decline in specific outcome measures. As such, progression is a concept that naturally lives on the time-axis. Modeling delay or slowing of progression will thus in general require nonlinear effects. The continuous-time extension (3) enables going from treatment effects that are additive parameters on the outcome scale to effects that modify the time course of disease relative to the placebo group. The most general such model, which we will refer to as the Time-PMRM, assumes that the mean outcome of the active treatment group at a given visit can be described as the mean outcome of the placebo group at another time point

$$g(t_{ij}; \boldsymbol{\alpha}, \boldsymbol{V}_{ij}\boldsymbol{\beta}) = f_0(\boldsymbol{V}_{ij}\boldsymbol{\beta} \cdot t_{ij}; \boldsymbol{\alpha}) \qquad (6)$$

where $\boldsymbol{\beta} = (1, \beta_1, \ldots, \beta_m)^\top$ and $\boldsymbol{V}_{ij} = (1, 0, \ldots, 0)$ if subject $i$ is in the placebo group, while for subject in the active arm $\boldsymbol{V}_{ij}$ is a vector of zeros except for the $(j+1)$th entry which is 1. Resultingly, for a subject in the active arm, the mean at visit $j > 0$ is $f_0(\beta_j \cdot t_{ij}; \boldsymbol{\alpha})$.

Figure 4 illustrates the relation between placebo mean values across visits that are modeled by the parameters $\boldsymbol{\alpha}$, the associated placebo trajectory $f_0$, and the mean values across visits of the active arm that are related to the placebo trajectory through the parameter vector $\boldsymbol{\beta}$.

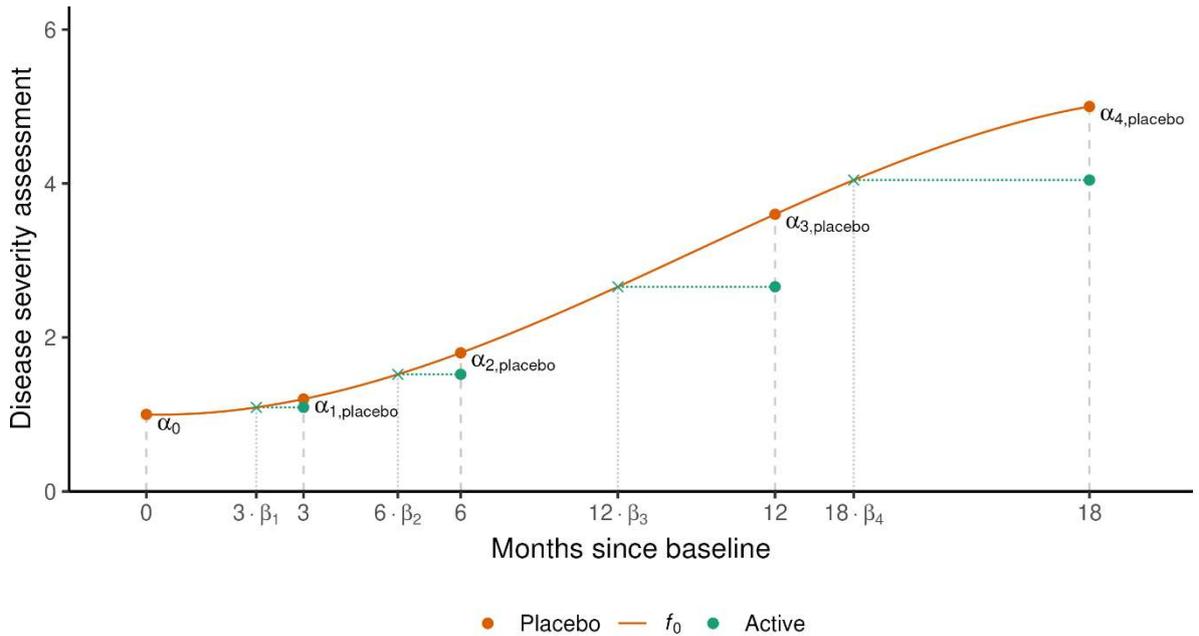

**Figure 4.** Illustration of the Time-PMRM for a hypothetical 18-month study. The figure shows

the relations between placebo mean values across visits that are modeled by the parameter vector **α**, the associated placebo trajectory $f_0$, and the mean values across visits of the active arm that are related to the placebo trajectory through the parameter vector **β** that describe the slowing of progression on the time-axis.

With the parametrization of the mean function in model (6), one gets a model that is in very similar to the MMRM by having one parameter per visit and treatment arm to describe the mean. However, there are four key differences compared to the MMRM:

- Mean values at post-baseline visits in the active arm are assumed to correspond to time points along the placebo mean trajectory; consequently, the model implicitly assumes that the active group mean cannot improve or decline beyond the range of the placebo group
- The parameter $\beta_j$ that describes the treatment effect at visit $j$ represents an effect of time and not the outcome measure; an estimate of 0.8 would represent a cumulative 20% slowing of progression at visit $j$
- Inference in the model is potentially based on the full shape of the placebo trajectory, so the estimated treatment effect at a given visit will potentially depend on data from more than one visit
- If treatment is expected to slow progression of disease and thus symptom manifestation across multiple measures, the model (4) can seamlessly be extended to multidimensional outcomes (e.g. one mean function $f_0^l$ for each outcome measure $l$) without increasing the number of parameters describing the treatment effect; $\beta_j$ models the unifying slowing of progression at visit $j$ across all outcomes

**Modeling proportional slowing of disease progression**

Similarly to how one can impose proportionality or other types of constraints on the cLDA model to develop low-dimensional parametrizations that efficiently represents certain types of treatment effects, one can do the same with the Time-PMRM.

The time-based counterpart to proportional reduction in decline (5) is a proportional slowing of progression (Figure 2D). This type of treatment effect can be modeled based on (6) using the same proportionality modeling of parameters as in (5), namely $\boldsymbol{\beta} = (1, \beta)^\top$ and

$$V_{ij} = \begin{cases} (1, 0) & \text{for treatment}(i) = \text{placebo} \\ (0, 1) & \text{for treatment}(i) = \text{active}. \end{cases}$$

This leads to the following model for subjects in the active arm

$$y_{ij} = f_0(\beta \cdot t_{ij}; \boldsymbol{\alpha}) + e_{ij}. \tag{7}$$

One would typically expect that $0 \leq \beta < 1$ which would indicate slowing of the disease time course, whereas $\beta_{\text{active}} < 0$ in principle would indicate reversal and $\beta_{\text{active}} > 1$ would indicate acceleration of the disease process. The relation between (horizontal) proportional slowing in this model and the (vertical) proportional reduction in decline in model (5) is akin to the relation between covariate effects in accelerated failure time models and proportional hazard models used in survival analysis.[15]

**Modeling delay of disease progression**

As opposed to slowing which represents a multiplicative factor of time, a delay is additive on the time scale and shares the unit of the time variable. A general time-delay PMRM is

$$g(t_{ij}; \boldsymbol{\alpha}, \boldsymbol{V}_{ij}\boldsymbol{\beta}) = f_0(t_{ij} - \boldsymbol{V}_{ij}\boldsymbol{\beta}; \boldsymbol{\alpha}) \qquad (8)$$

where $\boldsymbol{\beta} = (\beta_1, \ldots, \beta_m)^\top$ and $\boldsymbol{V}_{ij} = (0, \ldots, 0)$ for the baseline visit $j = 0$ or if subject $i$ is in the placebo group, and for subject in the active arm $\boldsymbol{V}_{ij}$ is a vector of zeros except for the $j$th entry which is 1. This model is a reparametrization of the mean function of the Time-PMRM (6). In model (8), the estimated $\beta$-parameters share the unit of the time variable instead of representing a proportionality factor of time.

A low-dimensional variant of the general delay model (8) is a constant delay model where the delay is constant over post-baseline visits $\boldsymbol{\beta} = \beta$ and $\boldsymbol{V}_{ij} = 1$ for post-baseline visits for subjects in the active arm and $\boldsymbol{V}_{ij} = 0$ otherwise.

A more realistic low-dimensional variant of (8) is a model that describes the maximal delay $\beta_{\max}$ which is achieved and maintained after visit $j'$ with a low-dimensional parametrization to model the delay up to this point (Figure 2D). A two-parameter model of this type can be achieved by using $\boldsymbol{\beta} = (\beta_1, \beta_{\max})^\top$ and design matrix

$$\boldsymbol{V}_{ij} = \begin{cases} \begin{pmatrix} 0 & 0 \\ 0 & 0 \end{pmatrix} & \text{for treatment}(i) = \text{placebo} \\ \begin{pmatrix} 1_{0<j<j'} & 0 \\ 0 & 1_{j \geq j'} \end{pmatrix} & \text{for treatment}(i) = \text{active} \end{cases}$$

where $1_{0<j<j'}$ is the indicator function that is 1 if $0 < j < j'$ and 0 otherwise.

# Case studies

In this section, we consider three case studies where we evaluate PMRMs in different trial scenarios. In the first case study, we consider fully simulated data of a 36-month trial in prodromal Alzheimer's disease. In the second case study, we consider resampled historical placebo-arm data from two 18-month trials in patients with mild Alzheimer's disease dementia. The purpose of the second case study is to investigate how the models perform in a more realistic scenario with associated potential misspecification of the error structure, realistic drop-out and a shorter trial duration. Furthermore, two types of artificially induced treatment effects are considered in the second case-study to explore the effects of potential misspecification of the treatment effect on the patient level. In the final case study, we give two examples of how to include covariates in the PMRM models.

In the two first case studies, six quantifications of treatment effects were compared. The first three were based on the cLDA model. First, treatment effects were evaluated as the contrasts between active and placebo at the final visit. Secondly, the treatment effects were evaluated as the difference in area under the curve (computed using the trapezoidal rule of model estimates). Thirdly, the global effect of treatment was tested using a type III F-test (no effect of treatment across visits). In the three last quantifications, the treatment effects were evaluated using the proportional decline PMRM (5), the Time-PMRM (6), and the proportional slowing Time-PMRM (7). PMRM treatment effects were tested using Wald tests (at last visit for the Time-PMRM). In all scenarios, the function $f_0$ describing the mean trajectory of the placebo arm was modeled as a natural cubic spline with knots at all visits. All models were implemented in R 4.0.4.[16] The cLDA model was implemented using the lmer function in the lme4 package.[17] Type III tests were done using the lmerTest package testing.[18] PMRMs were implemented as nonlinear mixed effects models using the gnls function in the nlme package[19] with custom mean functions used for the three different PMRMs. All models modeled correlations over time by assuming an unstructured covariance matrix common to all patients and all models were fitted using maximum likelihood estimation. R code for fitting the described PMRMs is available at online.[20]

## Case study 1: Simulated 36-month clinical trials in patients with prodromal Alzheimer's disease

To simulate a realistic 36-month clinical trial in prodromal Alzheimer's disease, patients were selected from the Alzheimer's disease neuroimaging initiative.[21] The following inclusion criteria were set up: at the baseline visit, patients must be diagnosed as having mild cognitive impairment, score less than or equal to 28 on the Mini Mental State Examination (MMSE, range 30-0, higher scores indicate less impairment)[22] and be amyloid positive according to a brain positron emission tomography scan or analysis of cerebrospinal fluid. 556 individuals with prodromal Alzheimer's disease met these inclusion criteria. The 13-item cognitive subscale of the Alzheimer's disease Assessment Scale (ADAS-cog, range 0-85, lower scores indicate less impairment)[23] was considered the primary outcome. Data from the baseline visits as well as visits at 6, 12, 18, 24 and 36 months after baseline were included. ADAS-cog scores were analyzed using a constrained longitudinal data analysis model with an unstructured covariance matrix and the estimated mean values across the six visits were

$$(19.6, 20.5, 20.9, 22.7, 23.8, 27.4)$$

while the estimated covariance matrix that describes the variation was

$$\begin{pmatrix} 45.1 & 40.0 & 45.1 & 54.9 & 53.6 & 60.8 \\ 40.0 & 57.8 & 54.4 & 66.3 & 64.1 & 74.7 \\ 45.1 & 54.4 & 72.0 & 80.0 & 77.6 & 93.1 \\ 54.9 & 66.3 & 80.0 & 109.8 & 99.3 & 121.7 \\ 53.6 & 64.1 & 77.6 & 99.3 & 111.4 & 127.8 \\ 60.8 & 74.7 & 93.1 & 121.7 & 127.8 & 191.4 \end{pmatrix}.$$

These estimates were used to simulate patient-level data. Each simulation consisted of a trial with either 300 or 500 patients per arm. Patient-level trajectories for patients in the placebo arm were simulated based on the above estimate of the mean across visits, while patients in the active arm had modified mean values across the visits depending on the type of treatment effect. Variants of the six types of treatment effects illustrated in Figure 2 were considered. The actual mean trajectories for the placebo arm and each of these six treatment effect scenarios are shown in Figure 5 and the equations for generating the treatment effects are given in the appendix. In the null scenarios, 5000 trials were simulated and in each of the scenarios with treatment effects, 1000 trials were simulated.

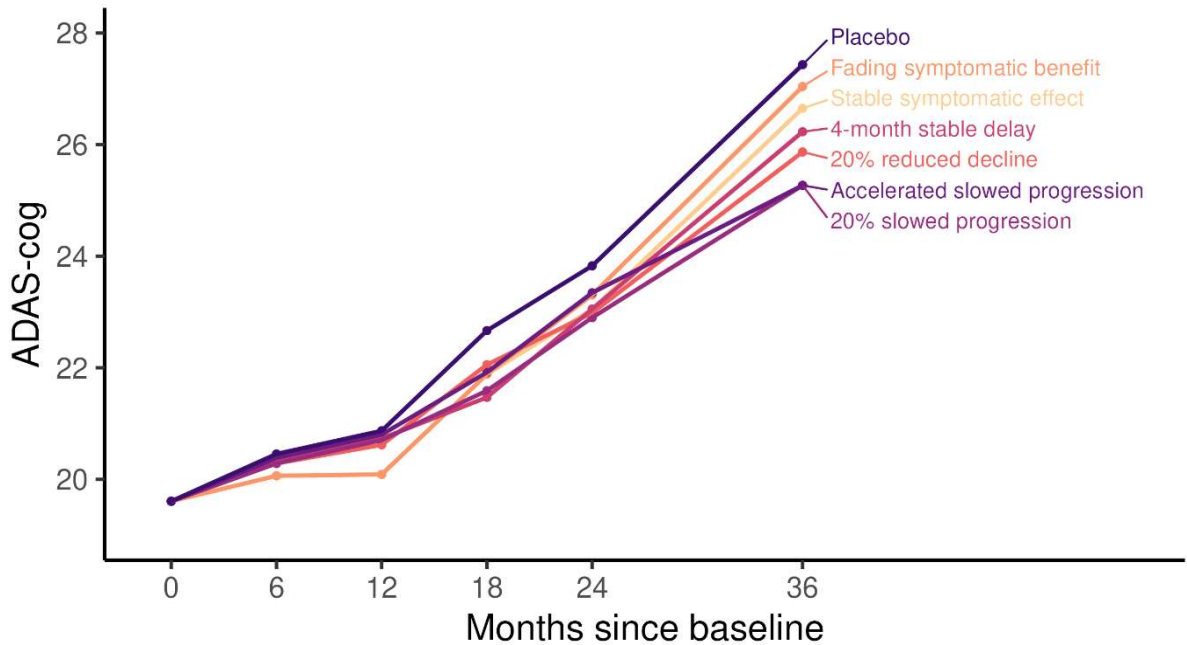

**Figure 5.** ADAS-cog placebo trajectory and the six considered treatment trajectories in case study 1.

**Type 1 error analysis**

When using uncalibrated significance cut-offs, the type 1 errors of the proportional decline PMRM and slowed progression Time-PMRM were inflated (0.037 and 0.066, respectively, with 300 patients per arm, 0.035 and 0.053, respectively, with 500 patients per arm; Table A1), while the type 1 errors for the other tests, including the Time-PMRM, were reasonably controlled. In the 22,000 simulated trials, only a single case of non-convergence of the Time-PMRM was observed. This simulation was excluded from the results.

**Power to detect treatment effects**

The power to detect treatment effects of the different models across simulations are given in Table 1. Significance cut-offs were recalibrated to achieve a one-sided type 1 error of 0.025 (except for the type III test that used a two-sided type 1 error of 0.05). The recalibration was done by using the 2.5% percentile of one-sided p-values of the different methods under the null scenarios as significance cut-offs. The results in Table 1 show that the cLDA treatment difference estimate at 36 months and the Time-PMRM slowing at 36 months had highly similar performance across the effect scenarios. In most scenarios, the type III test performed on par with or better than the 36-month treatment difference, however, this test answers the question "is there any treatment effect?" without offering a particularly interpretable quantification (i.e. a significant result could indicate a negative difference at one visit and a positive difference at a subsequent visit). Both the proportional decline PMRM and the slowing progression PMRM had poor power to detect treatment effects in the symptomatic effect scenarios but offered substantial power gains compared to the other methods in the scenarios where treatment effects increased over time.

**Effect estimates and coverage probability**

The proportional decline PMRM and proportional slowing Time-PMRM are misspecified in most of the simulation scenarios, but for the remaining models, a true effect under the model quantification can be derived across all scenarios. The true effects (when they can be derived) and the mean and median estimated effects across scenarios are given in Table A3 in the appendix. No indication of bias was found across methods and scenarios.

The distribution of maximum likelihood estimates from the methods across null scenarios (Figure A1 in appendix) suggested that while estimates were unbiased, some skewness was present in the distribution of estimates for the time-PMRM models. This skewness is a result of the natural cubic spline parametrization which gives greater variance when the active arm is matched to an extrapolation outside of the observed placebo trajectory (acceleration of disease).

The coverage probabilities of nominal two-sided 95% confidence intervals for the PMRM models are given in Table A4 in the appendix. Coverage probabilities are only given for the scenarios where a true effect quantification could be established. Across scenarios, the proportional decline PMRM and the time-PMRM all achieve coverage probabilities around or above 95%. The proportional slowing Time-PMRM achieved coverage probabilities between 86% and 92% in the four scenarios where a true effect quantification could be derived for this method.

**Evaluating the proportional slowing assumption**

To evaluate violations of the proportional slowing assumption, one can do a likelihood ratio test against the Time-PMRM. The proportional slowing assumption was found to be significantly worse than the Time-PMRM assumption (chi-square p-value < 0.05) in 5%/5% (300/500 patients per arm) of the *No effect* simulations, 27%/42% and 32%/50% of the *Stable symptomatic benefit* and *Fading symptomatic benefit* simulations, 14%/17% of the *20% reduced decline* scenarios, 31%/51% of the *4 months stable delay* scenarios, 9%/10% of the 20% *slowed progression* scenarios and 8%/14% of the *increasing slowing of progression* scenarios. Histograms of p-values are shown in figures A2 and A3 in the appendix.

## Case study 2: Clinical trials with simulated treatment effects based on historical trials in patients with mild Alzheimer's disease dementia

Pseudonymized patient-level data from the placebo arms of two 80-week trials in patients with mild to moderate Alzheimer's disease dementia was used for this second case study. The data were obtained through the TransCelerate Historical Trial Data Sharing Initiative.[24]

The primary outcome measure of this case study was the Clinical Dementia Rating (CDR) Sum of Boxes[25] (range 0-18, lower scores indicate less impairment) which was assessed at weeks 0 (baseline), 28, 52, and 80.

Data from 1024 patients were used for the simulations. All patients had a valid baseline assessment of CDR sum of boxes, 889 had valid assessments at week 28, 833 at week 52 and 769 at week 80. Figure 6 shows the patient-level trajectories.

Each simulation consisted of a trial with either 500 or 750 patients per arm. In the null scenarios, 5000 trials were simulated by randomly drawing patients from the full pool of patients with replacement. In each of the scenarios with treatment effects, 1000 trials were simulated similarly to the null scenario, but with treatment effects induced as described below.

We considered the four non-null treatment effect scenarios illustrated in Figure 7. In the first (null) scenario, there was no difference between placebo and active treatment. In the second scenario, a stable additive post-baseline benefit was seen in the active arm that amounted to a 20% reduced decline at the final visit. In the third scenario, there was a proportional reduction in decline of 20% in the active arm over all visits. In the fourth scenario, progression was delayed 16 weeks in the active arm, which built linearly on the time scale and was fully established after 40 weeks. In the final scenario, there was a 20% proportional slowing of progression in the active arm. While the models estimate population-level effects, the three latter types of effects can differ depending on whether they manifest on the mean or the patient level. The mean-level effects were computed based on the observed pointwise mean change from baseline scores at visits (thus excluding patients with no post-baseline observations from contributing to the effect estimate). These effects were added to the clinical scores of the patient trajectories sampled in the active arm. The patient-level effects were directly implemented on the individual patient level, thus allowing patients to have different benefits based on the observed progression of the sampled patient. We will refer to these implementations as *Mean* and *Subject-level* implementations of treatment effects. The details of the computations are given in Appendix A.

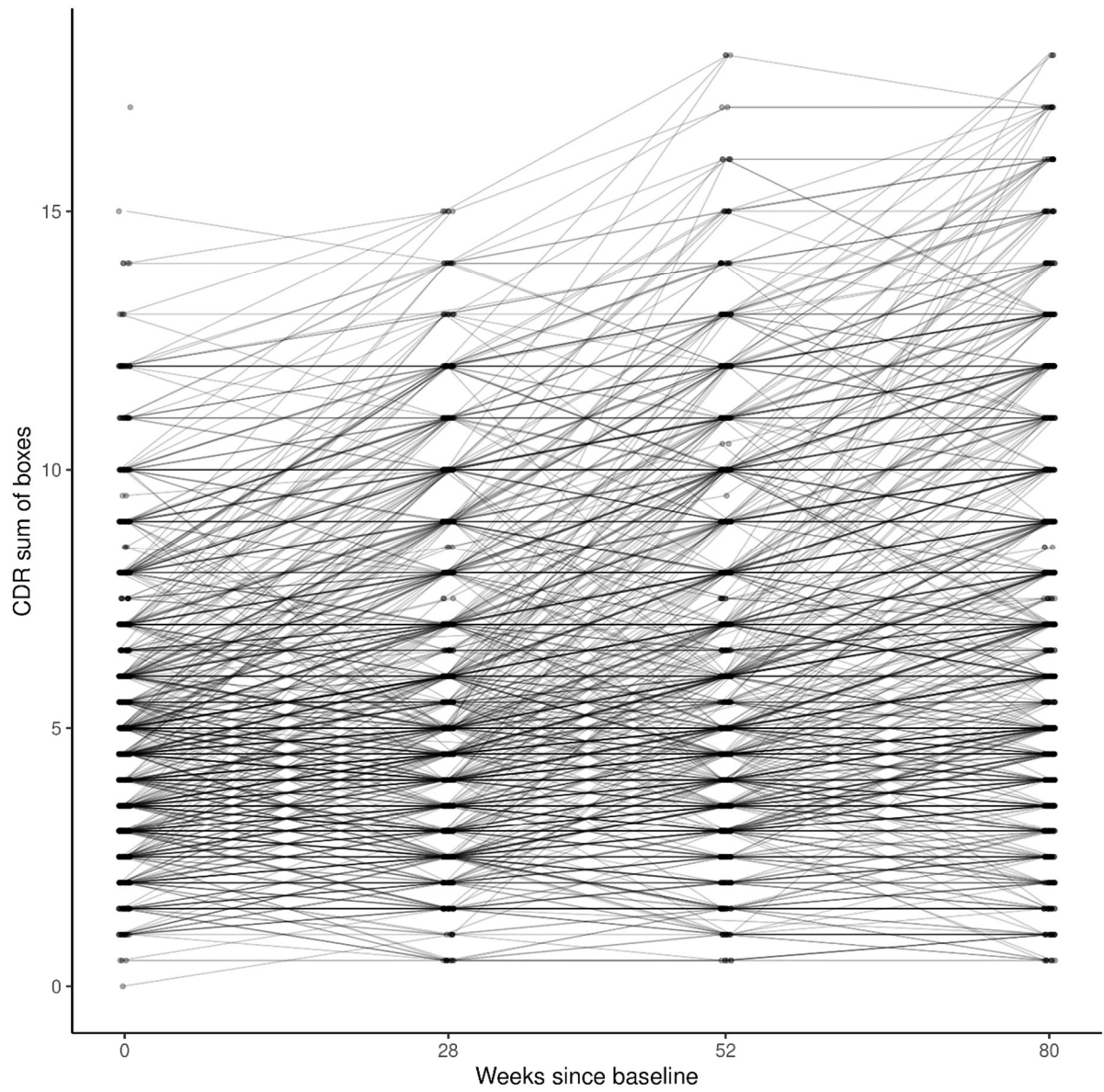

**Figure 6.** Patient-level trajectories of CDR sum of boxes for the 1024 patients used in case study 2.

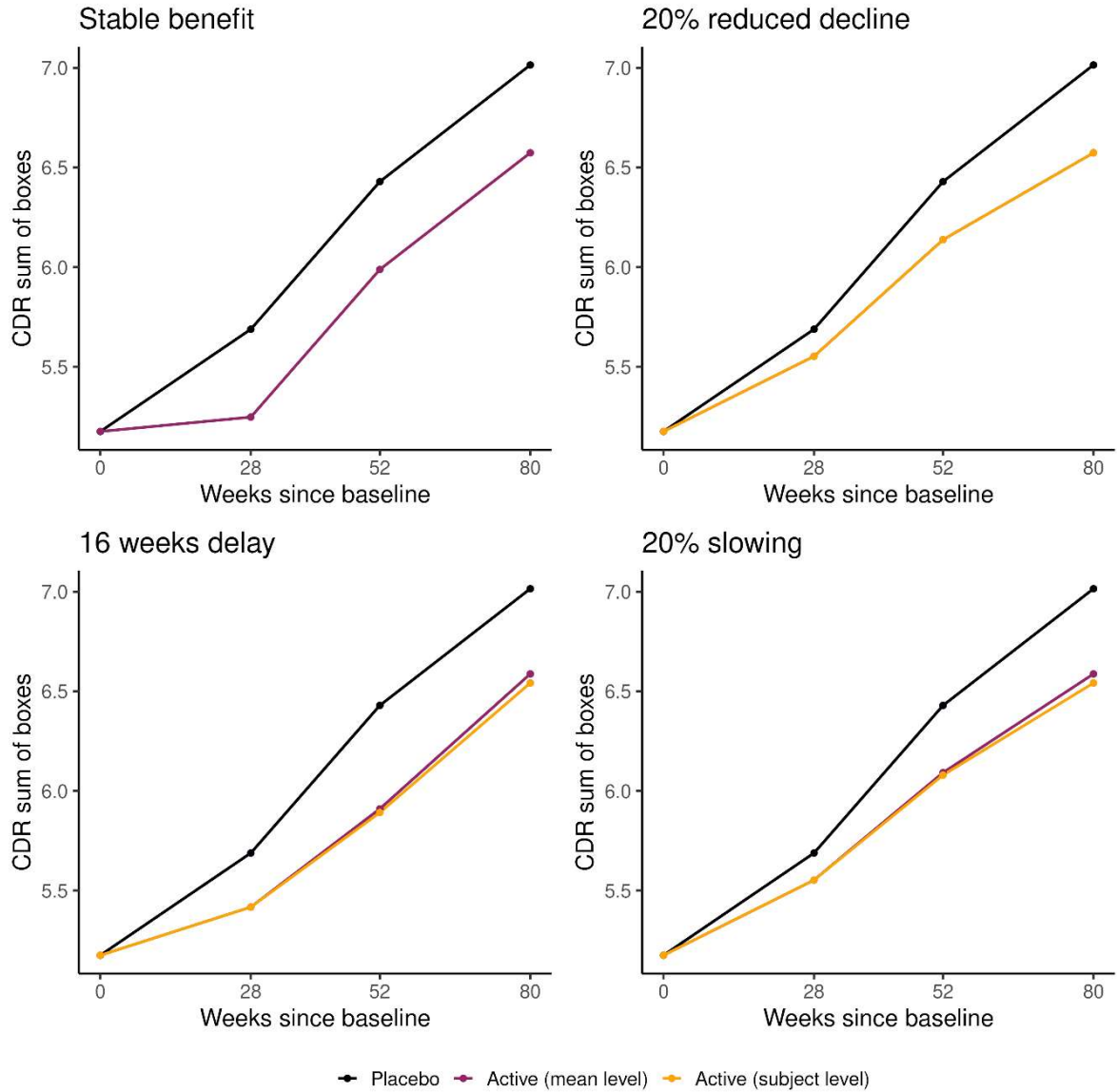

**Figure 7.** Mean trajectories of CDR sum of boxes across the scenarios considered. Points represent the pointwise mean values of CDR sum of boxes for the total population of 1024 patients. Note that no subject-level effect is implemented for the stable benefit scenario. For the 20% reduced decline scenario, the mean and subject-level effects perfectly coincide on the mean-trajectory level but not on the subject level. For the 16 weeks delay and 20% slowing, there are subtle differences between the effect implementations on the mean-trajectory level due to the nonlinear nature of these treatment effects.

The power to detect treatment effects of the different models across simulations are given in Table 2. Significance cut-offs were recalibrated to achieve a one-sided type 1 error of 0.025 (except for the type III test that used a two-sided type 1 error of 0.05) similarly to the previous case study. The type 1 errors of the PMRMs were inflated ranging from 0.035 to 0.052 when using uncalibrated significance cut-offs (Table A2).

**Power to detect treatment effects**

The results in Table 2 show that the cLDA treatment difference estimate at 80 weeks and the Time-PMRM slowing at 80 weeks again had highly similar performance across the effect scenarios. In most scenarios, the AUC difference and type III test performed on par with or better than the 80-week treatment difference. On average across all scenarios, the proportional decline PMRM and the slowing progression PMRM both had greater power than cLDA model and Time-PMRM, but the power gain associated with these lower-dimensional parametrizations of the treatment effect was considerably smaller than in case study 1. One possible explanation for this is that only 3 post-baseline assessments were included this case study.

## Case study 3: Including covariates in PMRMs

Covariate adjustment is fundamental for explaining variation in observed in data, and while PMRMs allow for adjustment of additive covariate terms such as covariate-by-categorical visit just as in linear mixed-effects models, PMRMs also offer the possibility adjusting for covariates that have nonlinear effects. In this case study, we will consider two examples of covariate adjustment.

**Initial improvements in both arms**

Initial improvements in clinical assessments are often observed in clinical trials, and may be ascribed to many factors, including placebo effects, training effects (e.g. by repeated exposed to the same type of cognitive scales), improved standard of care, or regression to the mean.[26] The time-PMRM models relate the visit mean outcome values of the active arm horizontally to the estimated trajectory of the placebo arm $f_0$, and initial improvements may thus be discordant with the interpretation of slowing or delay of disease difficult.

Figure 8 shows mean trajectories from a hypothetical 78-week clinical trial where the active treatment slows progression of disease proportionally by 30% compared to placebo. In addition to this slowing of disease, an initial improvement is present at visits 12 and 26 weeks post baseline. The initial improvement is identical in both arms.

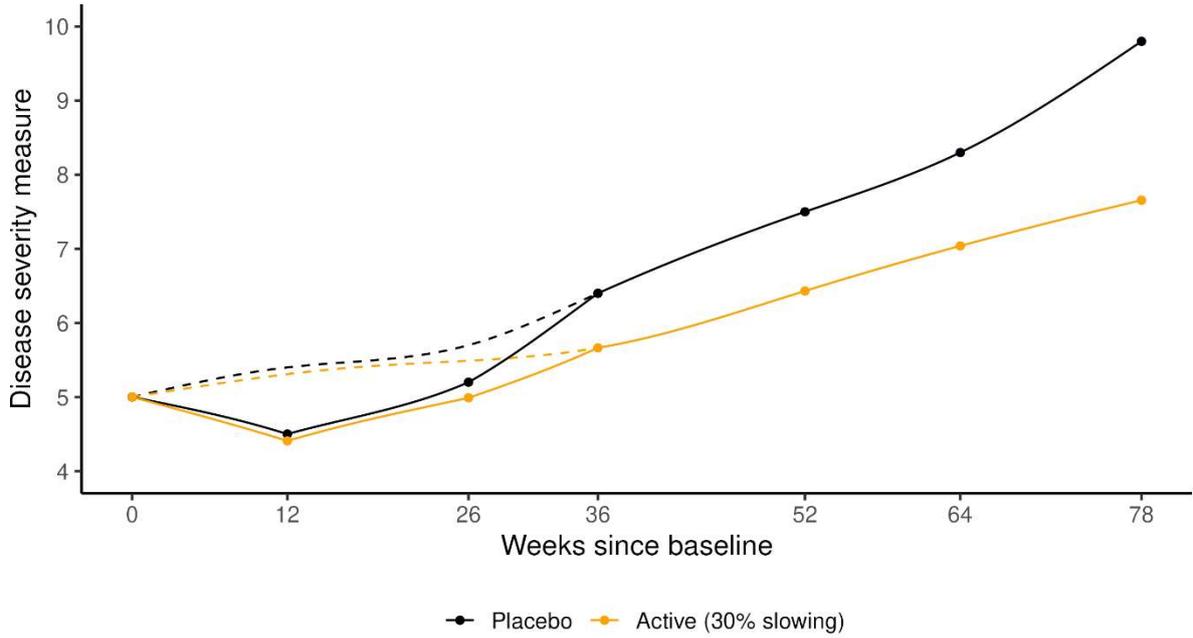

**Figure 8.** Mean trajectories of a hypothetical 78-week clinical trial where the active treatment slows progression of disease proportionally by 30% compared to placebo. In addition to this slowing of disease, an initial similar improvement is present at visits 12 and 26 weeks post baseline. Dashed lines show the assumed natural history trajectory and full lines show the actual mean values.

Since the initial improvement is independent of treatment, the first two post-baseline visits are not consistent with the proportional slowing model (7). To model this improvement, one can include additive covariates to model (7) describing the improvement at these two visits by

$$y_{ij} = f_0(\beta \cdot t_{ij}; \boldsymbol{\alpha}) + \delta_{12} \cdot 1_{t_j=1} + \delta_{26} \cdot 1_{t_j=26} + e_{ij}$$

where $1_{t_{ij}=12}$ is the indicator function that is 1 when $t_{ij} = 12$ and 0 otherwise and $\delta_{12}$ it the associated parameter describing the deviation at this visit, and similarly for the week 26 parameters.

Figure 9 shows an example of maximum-likelihood estimates of the mean trajectories of the proportional slowing Time-PMRM (7) and the above extension with parameters modeling the initial improvement for a simulated example trial with 300 subjects per arm using the mean trajectories shown in Figure 8.

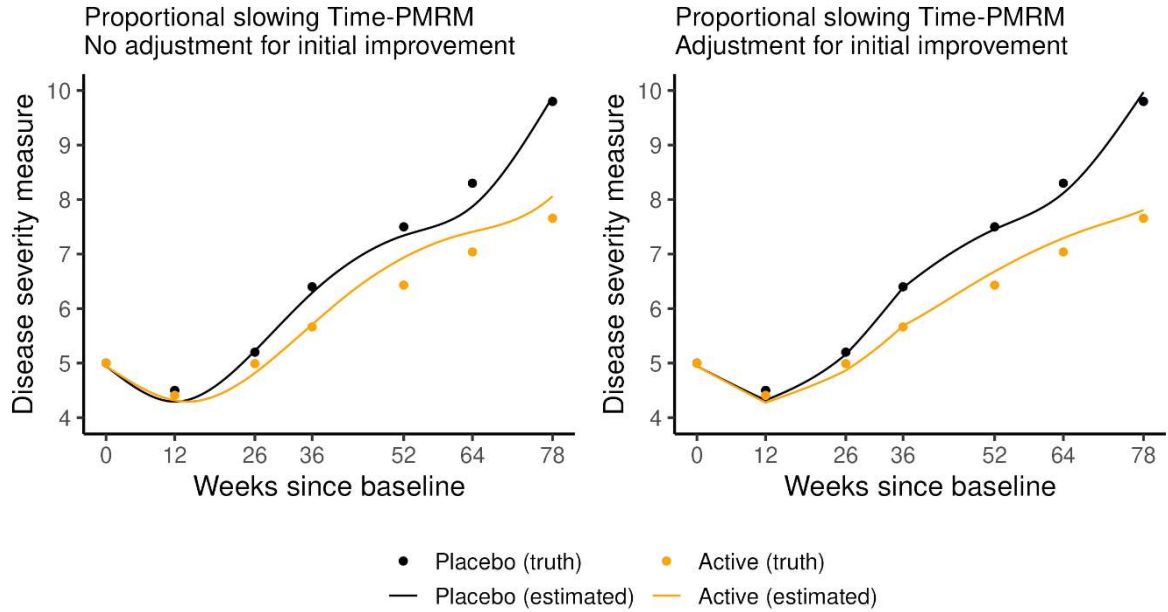

**Figure 9.** Maximum-likelihood estimates of mean trajectories of simulated trial data with initial improvement in both arms. Left figure shows estimates of the proportional slowing Time-PMRM model and right figure shows estimates of the same model with covariates modeling the initial improvement at 12 and 26 weeks. The trajectory of the initial improvement is displayed using linear interpolation.

**Subgroups with different progression rates**

Certain patient characteristics may be associated with increased progression rate, which can in turn be used for trial enrichment.[27] One of the ideas behind enrichment in trials of progressive diseases is that faster progressors may offer a better window to detect a benefit. If a treatment slows the progression of disease similarly across subgroups, groups of fast progressors will have a greater benefit.

Suppose we can define two groups of patients at baseline, and one has 10% faster progression rate. Figure 10 shows mean trajectories from such a hypothetical 78-week trial with active treatment resulting in a 30% proportional slowing of progression in both groups of patients. Even with an identical slowing of progression in the two groups, the nonlinear nature of the trajectory means that the effect manifests differently when measured as points difference on the outcome scale. At week 78, the average treatment difference is -1.97 points, but it is -2.23 in group 1 (50% of patients, faster progressors) and -1.72 in group 2 (remaining 50%).

In the proportional slowing model (7), such nonlinear interaction between treatment and subgroup membership can be modeled efficiently by including one additional parameter $\rho$ that describes the relation between the speed of progression in groups 1 and 2

$$y_{ij} = f_0\big(\beta \cdot (1 - \rho \cdot 1_{\text{group}(i)=2}) \cdot t_{ij}; \boldsymbol{\alpha}\big) + e_{ij}.$$

This model requires a total of 9 parameters to describe the mean structure modeling the nonlinear interaction between rate of progression and treatment effects. In contrast, a cLDA model would require 25 parameters to have sufficient flexibility to describe this treatment effect across visits.

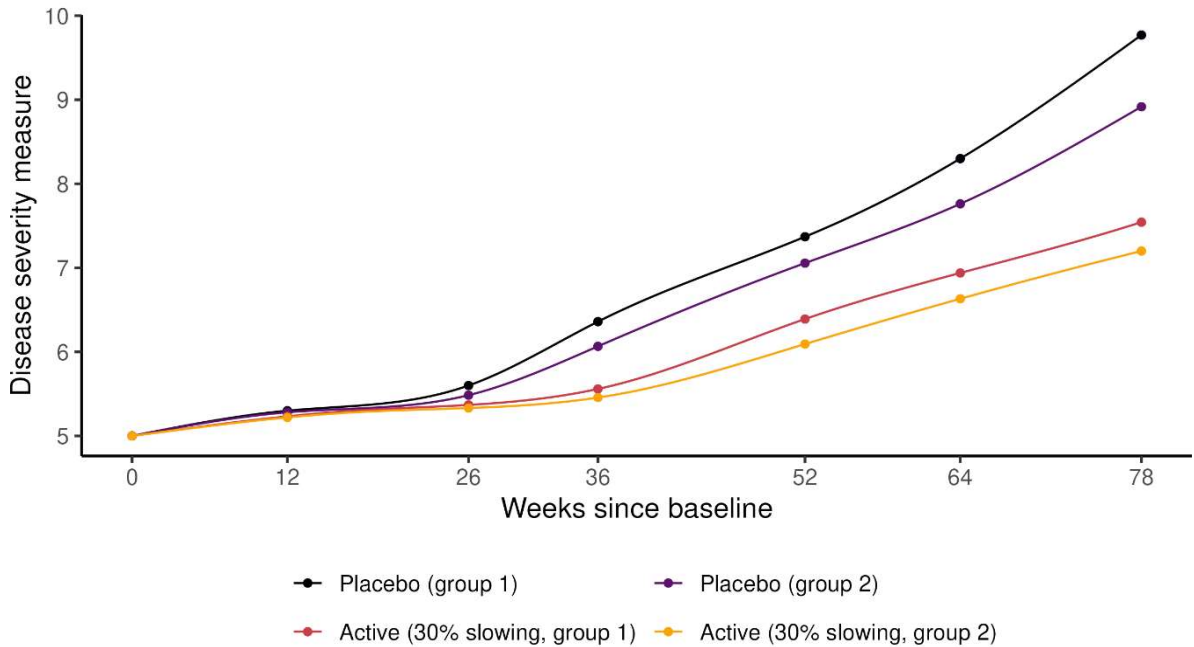

**Figure 10.** Hypothetical mean trajectories of a disease severity outcome in a 78-week trial where one subgroup had 10% faster progression. The treatment effect was 30% slowing of progression in both groups.

## Discussion

This paper highlighted the utility of the class of PMRMs that are continuous-time nonlinear models that enable estimation of non-standard treatment effects such as delay or slowing of disease progression. These quantifications of treatment effects may be more interpretable than conventional treatment differences.

In experiments based on both simulated and historical data from Alzheimer's disease clinical trials, the Time-PMRM was highly similar to conventionally used models in terms of power to detect treatment effects, but quantified treatment effects in terms of slowing of disease progression at a given visit instead of the difference at the outcome scale. Furthermore, PMRMs that modeled treatment effects as proportional reduction in decline or proportional slowing of progression on average had the greatest power to detect treatment effects in scenarios where the efficacy increased over time. The gain in efficiency was substantial in some cases. While alternative methods for quantifying and testing a treatment effect, such as the AUC difference or a type III test, may offer similar or better power in some scenarios, the associated treatment effects may not be particularly interpretable.

Compared to the conventional cLDA model, the PMRMs showed increased type 1 error rates in the two case studies. The inflated type 1 error was less pronounced for the general Time-PMRM than the two PMRMs with proportionality assumptions and seemed to be somewhat alleviated by increasing sample size. It is generally acknowledged in the literature that likelihood ratio and Wald tests in nonlinear mixed effects models can increase type 1 error rates.[28, 29] In the present examples, this was likely due in part to the estimation procedure and distributional deviations in the test statistics. In the second case study, an additional contributing factor was likely also that resampled historical observations were used, which led to some additional deviations from the assumed model (e.g. assumed normality of residual error terms). Accordingly, a potential type 1 error inflation is something that should be considered when considering the use of the PMRM framework in a clinical trial setting. One potential way to address this is by recalibrating the significance cut-off based on resampling techniques, similarly to what was done in this paper.

One may wish to estimate the types of treatment effects suggested here because they match prior knowledge about the effect of an intervention (e.g. that it is slowing or stopping progression of molecular-level disease processes). However, the Time-PMRM approach may also simply be used to quantify a treatment effect in terms of time which offers a readily interpretable and relevant measure of treatment effect compared to points on an outcome scale whose clinical meaningfulness may not be well established.

The time-based PMRMs have several other appealing properties. For example, the ability to quantify treatment effects in terms of time enables better comparison of effects across trials with different endpoints. With a straightforward extension, they also offer the possibility of using less rigid visit structures in clinical trials since treatment effects can be estimated based on the entirety of patient trajectories. Finally, the time-based PMRMs can be extended to allow efficient estimation of treatment effects using multiple different endpoints simultaneously, for example using a single parameter to model slowing of time-progression across multiple endpoints. Notably, if a treatment is slowing the progression of disease resulting in similar slowing across outcomes, the same level of model efficiency cannot be achieved in general by modeling treatment differences on the outcome scale (including proportional decline models). The reason for this is that decline in a given outcome measure is largely determined by properties such as its sensitivity to change in the trial population at a given time point, so one would in general expect this slowing to manifest as different levels of rescued decline across outcome scales.

In the present work, we mainly considered simplified models with a minimal fixed-effect design and no explicit random effects. While it is straightforward to introduce additive fixed and random effects similarly to a linear mixed-effects model, the PMRMs also allow specification of nonlinear fixed and random effects that can for example modify the time-progression of disease for a group or on the individual level. This adds a level of complexity, and one should consider on a case-by-case basis how to properly adjust for baseline covariates. One particularly relevant aspect to consider is the potential to include patient-level differences in disease progression. This has been a major focus in Alzheimer's disease, where a range of exploratory disease progression models with latent variables describing patient-level progression has been developed with a focus on predicting the disease stage and future decline of an individual patient.[30-35] Future work should address the most appropriate use of the PMRM methodology in relation to covariates, random effects and multivariate outcomes in clinical trials.

**Acknowledgements**

Development, Inc., AstraZeneca AB, Boehringer Ingelheim International GmbH, Bristol Myers Squibb Company, EMD Serono Research & Development Institute, Inc., Genentech, Inc., GlaxoSmithKline LLC, Eli Lilly and Co., Janssen Research & Development, LLC, Novartis Pharma AG, Novo Nordisk A/S, Pfizer Inc., Sanofi USA, Shionogi & Co., Ltd., and UCB Biosciences GmbH.

**Data Availability Statement**

The data that was used in this study are available from TransCelerate Biopharma Inc.'s Historical Trial Sharing project. Descriptions of how to access data are avaiable on TransCelerate's website https://www.transceleratebiopharmainc.com/initiatives/historical-trial-data-sharing/.

**Table 1.** Power to detect a treatment effect using different methods across different simulated scenarios. Significance cut-offs have been recalibrated to achieve a one-sided type 1 error of 0.025 (two-sided type 1 error of 0.05 for type III test). The corresponding unadjusted power estimates are found in Table A1 in the appendix. **Bold** numbers indicate the method with best power in each scenario.

| Patients per arm | Effect | cLDA (36 months)[a] | cLDA AUC difference[b] | cLDA type III test[c] | PMRM proportional decline | Time-PMRM (36 months)[d] | Time-PMRM proportional slowing |
|---|---|---|---|---|---|---|---|
| 300 | No effect | 0.025 | 0.025 | 0.05 | 0.025 | 0.025 | 0.025 |
| | Stable symptomatic benefit | 0.150 | **0.324** | 0.277 | 0.049 | 0.138 | 0.024 |
| | Fading symptomatic benefit | 0.069 | 0.212 | **0.318** | 0.012 | 0.059 | 0.016 |
| | 20% reduced decline | 0.457 | 0.300 | 0.303 | **0.502** | 0.453 | 0.285 |
| | 4 months stable delay of progression | 0.323 | 0.309 | **0.611** | 0.374 | 0.344 | 0.569 |
| | 20% slowed progression | 0.727 | 0.465 | 0.685 | 0.789 | 0.741 | **0.846** |
| | Increasing slowing of progression | 0.731 | 0.318 | 0.753 | 0.699 | 0.741 | **0.909** |
| 500 | No effect | 0.025 | 0.025 | 0.05* | 0.025 | 0.025 | 0.025 |
| | Stable symptomatic benefit | 0.229 | **0.509** | 0.454 | 0.057 | 0.201 | 0.027 |
| | Fading symptomatic benefit | 0.084 | 0.346 | **0.517** | 0.009 | 0.076 | 0.019 |
| | 20% reduced decline | 0.647 | 0.486 | 0.480 | **0.707** | 0.651 | 0.489 |
| | 4 months stable delay of progression | 0.458 | 0.504 | **0.883** | 0.475 | 0.477 | 0.777 |
| | 20% slowed progression | 0.917 | 0.719 | 0.918 | 0.952 | 0.920 | **0.977** |
| | Increasing slowing of progression | 0.916 | 0.449 | 0.942 | 0.859 | 0.924 | **0.984** |

[a] Mean difference between active and placebo at 36 months

[b] Mean difference in area under the curve between active and placebo from 0 to 36 months computed using the trapezoidal rule of estimated

[c] Difference between active and placebo throughout the study (test of treatment effect across all visits)

[d] Mean slowing of active arm relative to placebo at 36 months

**Table 2.** Power to detect a treatment effect using different methods across different simulated scenarios. Significance cut-offs have been recalibrated to achieve a one-sided type 1 error of 0.025 (two-sided type 1 error of 0.05 for type III test). The corresponding unadjusted power estimates are found in Table A2 in the appendix. **Bold** numbers indicate the method with best power in each scenario.

| Patients per arm | Effect | Effect implementation | cLDA (80 weeks)[a] | cLDA AUC difference[b] | cLDA type III test[c] | PMRM proportional decline | Time-PMRM (80 weeks)[d] | Time-PMRM proportional slowing |
|---|---|---|---|---|---|---|---|---|
| 500 | No effect | — | 0.025 | 0.025 | 0.05 | 0.025 | 0.025 | 0.025 |
| | Stable benefit | Mean | 0.503 | 0.911 | **0.926** | 0.481 | 0.528 | 0.642 |
| | 20% reduced decline | Mean | 0.503 | 0.505 | 0.403 | **0.539** | 0.528 | 0.455 |
| | | Subject | 0.629 | 0.610 | 0.508 | **0.662** | 0.634 | 0.553 |
| | 16 weeks delayed progression | Mean | 0.485 | **0.859** | 0.835 | 0.564 | 0.503 | 0.786 |
| | | Subject | 0.731 | **0.958** | 0.938 | 0.813 | 0.752 | 0.911 |
| | 20% slowed progression | Mean | 0.485 | **0.551** | 0.443 | 0.542 | 0.503 | 0.543 |
| | | Subject | 0.704 | 0.687 | 0.597 | 0.736 | 0.716 | **0.733** |
| 750 | No effect | — | 0.025 | 0.025 | 0.05 | 0.025 | 0.025 | 0.025 |
| | Stable benefit | Mean | 0.691 | 0.976 | **0.986** | 0.621 | 0.692 | 0.832 |
| | 20% reduced decline | Mean | 0.691 | 0.650 | 0.533 | **0.700** | 0.692 | 0.650 |
| | | Subject | **0.818** | 0.754 | 0.657 | **0.818** | 0.816 | 0.759 |
| | 16 weeks delayed progression | Mean | 0.670 | **0.948** | 0.940 | 0.719 | 0.663 | 0.898 |
| | | Subject | 0.881 | **0.992** | 0.990 | 0.924 | 0.888 | 0.970 |
| | 20% slowed progression | Mean | 0.670 | 0.702 | 0.571 | 0.693 | 0.663 | **0.740** |
| | | Subject | 0.864 | 0.820 | 0.758 | 0.868 | 0.864 | **0.891** |

[a]Mean difference between active and placebo at 80 weeks

[b]Mean difference in area under the curve between active and placebo from 0 to 80 months computed using the trapezoidal rule of estimated

[c]Difference between active and placebo throughout the study (test of treatment effect across all visits)

[d]Mean slowing of active arm relative to placebo at 80 weeks

# Appendix: Simulation study

# Case study 1: Simulation of treatment effects

Data in case study 1 were simulated from the following model

$$y_{ij} = \alpha_{j,\text{treatment}(i)} + e_{ij}$$

where

$$\boldsymbol{\alpha}_{\text{placebo}} = \left(\alpha_{j,\text{placebo}}\right)_{j=0,\ldots,5} = (19.6,\ 20.5,\ 20.9,\ 22.7,\ 23.8,\ 27.4)$$

and

$$(e_{ij})_{j=0,\ldots,5} \sim N_6(0, \boldsymbol{R})$$

with

$$\boldsymbol{R} = \begin{pmatrix} 45.1 & 40.0 & 45.1 & 54.9 & 53.6 & 60.8 \\ 40.0 & 57.8 & 54.4 & 66.3 & 64.1 & 74.7 \\ 45.1 & 54.4 & 72.0 & 80.0 & 77.6 & 93.1 \\ 54.9 & 66.3 & 80.0 & 109.8 & 99.3 & 121.7 \\ 53.6 & 64.1 & 77.6 & 99.3 & 111.4 & 127.8 \\ 60.8 & 74.7 & 93.1 & 121.7 & 127.8 & 191.4 \end{pmatrix}.$$

Let $\vartheta(t)$ denote the linear interpolation at time point $t$ of the mean values $\boldsymbol{\alpha}_{\text{placebo}}$ at the associated visit times ($t_j = 0,\ 6,\ 12,\ 18,\ 24$ and $36$ months since baseline). Treatment effects in the different simulated active arm scenarios were generated as follows.

**Stable symptomatic benefit**

$$\left(\alpha_{j,\text{active}}\right)_{j=0,\ldots,5} = \boldsymbol{\alpha}_{\text{placebo}} + (0,\ 0.39,\ 0.78,\ 0.78,\ 0.78,\ 0.78).$$

**Fading symptomatic benefit**

$$\left(\alpha_{j,\text{active}}\right)_{j=0,\ldots,5} = \boldsymbol{\alpha}_{\text{placebo}} + (0,\ 0.39,\ 0.78,\ 0.78,\ 0.52,\ 0.39).$$

**20% reduced decline**

$$\left(\alpha_{j,\text{active}}\right)_{j=0,\ldots,5} = 0.8 \cdot \left(\boldsymbol{\alpha}_{\text{placebo}} - \alpha_{0,\text{placebo}}\right) + \alpha_{0,\text{placebo}}.$$

**4 months stable delay of progression**

$$\left(\alpha_{j,\text{active}}\right)_{j=0,\ldots,5} = \left(\vartheta(0), \vartheta(6-1), \vartheta(12-2), \vartheta(18-4), \vartheta(24-4), \vartheta(36-4)\right).$$

**20% slowed progression**

$$\left(\alpha_{j,\text{active}}\right)_{j=0,\ldots,5} = \left(\vartheta(0.8 \cdot t_j)\right)_{j=0,\ldots,5}.$$

**Increasing slowing of progression**

$$\left(\alpha_{j,\text{active}}\right)_{j=0,\ldots,5} = \left(\vartheta(0), \vartheta(6-0.5), \vartheta(12-1), \vartheta(18-2.5), \vartheta(24-2.5), \vartheta(36-7.2)\right).$$

The mean values of the placebo arm as well as the six treatment arms are given in the table below

| Group | 0 months | 6 months | 12 months | 18 months | 24 months | 36 months |
|---|---|---|---|---|---|---|
| Placebo | 19.6 | 20.5 | 20.9 | 22.7 | 23.8 | 27.4 |
| Stable symptomatic benefit | 19.6 | 20.1 | 20.1 | 21.9 | 23.0 | 26.7 |
| Fading symptomatic benefit | 19.6 | 20.1 | 20.1 | 21.9 | 23.3 | 27.0 |
| 20% reduced decline | 19.6 | 20.3 | 20.6 | 22.1 | 23.0 | 25.9 |
| 4 months stable delay of progression | 19.6 | 20.3 | 20.7 | 21.5 | 23.1 | 26.2 |
| 20% slowed progression | 19.6 | 20.3 | 20.7 | 21.6 | 22.9 | 25.3 |
| Increasing slowing of progression | 19.6 | 20.4 | 20.8 | 21.9 | 23.3 | 25.3 |

## Case study 2: Simulation of treatment effects

Patient-level data in case study 2 were first sampled from the full pool of available patients, and if the patient was sampled in the active arm, the observed scores were altered to represent the treatment effect.

To generate the treatment effects, the pointwise mean change from baseline at each visit was estimated based on the following model using all observed data for subjects with post-baseline data

$$y_{ij} - y_{i0} = \delta_j + e_{ij}, \quad j > 0$$

where

$$(e_{ij})_{j=1,\ldots,3} \sim N_3(0, \mathbf{R}), \quad \mathbf{R} = \begin{pmatrix} \sigma_1^2 & 0 & 0 \\ 0 & \sigma_2^2 & 0 \\ 0 & 0 & \sigma_3^2 \end{pmatrix}.$$

The estimated mean change from baseline parameters were

$$\hat{\boldsymbol{\delta}} = (\hat{\delta}_j)_{j=1,\ldots,3} = (0.68, 1.46, 2.20).$$

For treatment effects that represented proportional reduction in decline, delay of progression or proportional slowing of disease, the effect on patient trajectories is implemented in two different ways. The details of how these are implemented are given below.

### Mean-level treatment effects

In the active arm, each patient was randomly sampled from the full population of available patients. Let $\vartheta(t)$ denote the linear interpolation at time point $t$ of the mean change from baseline values $\hat{\boldsymbol{\delta}}$ at the associated visit times ($t_j = 28, 52,$ and $80$ weeks since baseline). For each post-baseline visit $j$ with available data $y_{ij}$, the patient data for the active arm $\tilde{y}_{ij}$ were generated as follows.

**Stable symptomatic benefit**

$$\tilde{y}_{ij} = y_{ij} - 0.2 \cdot \hat{\delta}_3, \quad j > 0.$$

**20% reduced decline**

$$\tilde{y}_{ij} = y_{ij} - 0.2 \cdot \hat{\delta}_j, \quad j \geq 0.$$

**16 weeks delayed progression (linearly building to week 40)**

$$\tilde{y}_{ij} = y_{ij} - \left(\hat{\delta}_j - \vartheta(t_j - \min(t_j, 40)/40 \cdot 16)\right), \quad j > 0.$$

**20% slowed progression**

$$\tilde{y}_{ij} = y_{ij} - \left(\hat{\delta}_j - \vartheta(0.8 \cdot t_j)\right), \quad j > 0.$$

## Patient-level treatment effects

In the active arm, each patient was randomly sampled from the full population of available patients. Now, let $\vartheta_i(t)$ denote the linear interpolation at time point $t$ of the patient-level data $y_{ij}$ at the associated visit times ($t_j = 0, 28, 52,$ and 80 weeks since baseline). For each visit $j$ with available data $y_{ij}$, the patient data for the active arm $\tilde{y}_{ij}$ were generated as follows.

**20% reduced decline**

$$\tilde{y}_{ij} = 0.8 \cdot y_{ij}, \quad j \geq 0.$$

**16 weeks delayed progression (linearly building to week 40)**

$$\tilde{y}_{ij} = \vartheta_i(t_j - \min(t_j, 40)/40 \cdot 16), \quad j \geq 0.$$

**20% slowed progression**

$$\tilde{y}_{ij} = \vartheta_i(0.8 \cdot t_j), \quad j \geq 0.$$

The mean values of the placebo arm as well as the seven treatment arms are given in the table below

| Group | 0 weeks | 28 weeks | 52 weeks | 80 weeks |
|---|---|---|---|---|
| Placebo | 5.18 | 5.69 | 6.43 | 7.01 |
| Stable symptomatic benefit (mean level) | 5.18 | 5.25 | 5.99 | 6.57 |
| 20% reduced decline (mean level) | 5.18 | 5.55 | 6.14 | 6.57 |
| 16 weeks delayed progression (mean level) | 5.18 | 5.42 | 5.91 | 6.59 |
| 20% slowed progression (mean level) | 5.18 | 5.55 | 6.09 | 6.59 |
| **20% reduced decline (patient level)** | 5.18 | 5.55 | 6.14 | 6.57 |
| **16 weeks delayed progression (patient level)** | 5.18 | 5.42 | 5.89 | 6.54 |
| **20% slowed progression (patient level)** | 5.18 | 5.55 | 6.08 | 6.54 |

# Power simulations without recalibration of significance level

**Table A1.** Power to detect a treatment effect using different methods across different simulated scenarios in Case Study 1 (one-sided testing at the 0.025 level except for type III test that used two-sided testing at the 0.05 level).

| Patients per arm | Effect | cLDA (36 months) | cLDA AUC difference | cLDA type III test | PMRM proportional decline | Time-PMRM (36 months) | Time-PMRM proportional slowing |
|---|---|---|---|---|---|---|---|
| 300 | No effect | 0.024 | 0.028 | 0.056 | 0.037 | 0.029 | 0.066 |
| | Stable symptomatic benefit | 0.147 | 0.349 | 0.286 | 0.067 | 0.146 | 0.074 |
| | Fading symptomatic benefit | 0.068 | 0.227 | 0.337 | 0.015 | 0.061 | 0.050 |
| | 20% reduced decline | 0.450 | 0.311 | 0.319 | 0.574 | 0.468 | 0.475 |
| | 4 months stable delay of progression | 0.318 | 0.326 | 0.637 | 0.420 | 0.354 | 0.704 |
| | 20% slowed progression | 0.725 | 0.491 | 0.699 | 0.838 | 0.756 | 0.932 |
| | Increasing slowing of progression | 0.726 | 0.330 | 0.767 | 0.755 | 0.757 | 0.956 |
| 500 | No effect | 0.023 | 0.23 | 0.49 | 0.035 | 0.029 | 0.053 |
| | Stable symptomatic benefit | 0.225 | 0.504 | 0.451 | 0.081 | 0.212 | 0.063 |
| | Fading symptomatic benefit | 0.081 | 0.334 | 0.514 | 0.012 | 0.081 | 0.035 |
| | 20% reduced decline | 0.638 | 0.474 | 0.478 | 0.773 | 0.665 | 0.633 |
| | 4 months stable delay of progression | 0.450 | 0.495 | 0.882 | 0.548 | 0.492 | 0.860 |
| | 20% slowed progression | 0.913 | 0.710 | 0.918 | 0.960 | 0.925 | 0.992 |
| | Increasing slowing of progression | 0.912 | 0.441 | 0.942 | 0.901 | 0.933 | 0.991 |

**Table A2.** Power to detect a treatment effect using different methods across different simulated scenarios in Case Study 2 (one-sided testing at the 0.025 level except for type III test that used two-sided testing at the 0.05 level).

| Patients per arm | Effect | Effect implementation | cLDA (80 weeks) | cLDA AUC difference | cLDA type III test | PMRM proportional decline | Time-PMRM (80 weeks) | Time-PMRM proportional slowing |
|---|---|---|---|---|---|---|---|---|
| 500 | No effect | — | 0.031 | 0.026 | 0.048 | 0.042 | 0.043 | 0.052 |
| | Stable benefit | Mean | 0.540 | 0.916 | 0.924 | 0.549 | 0.619 | 0.773 |
| | 20% reduced decline | Mean | 0.540 | 0.519 | 0.390 | 0.627 | 0.619 | 0.601 |
| | | Subject | 0.662 | 0.628 | 0.501 | 0.734 | 0.733 | 0.707 |
| | 16 weeks delayed progression | Mean | 0.509 | 0.866 | 0.831 | 0.641 | 0.593 | 0.850 |
| | | Subject | 0.764 | 0.962 | 0.936 | 0.864 | 0.817 | 0.936 |
| | 20% slowed progression | Mean | 0.509 | 0.565 | 0.432 | 0.630 | 0.593 | 0.677 |
| | | Subject | 0.729 | 0.708 | 0.585 | 0.800 | 0.795 | 0.841 |
| 750 | No effect | — | 0.024 | 0.027 | 0.057 | 0.035 | 0.035 | 0.042 |
| | Stable benefit | Mean | 0.684 | 0.981 | 0.987 | 0.675 | 0.746 | 0.892 |
| | 20% reduced decline | Mean | 0.684 | 0.668 | 0.566 | 0.756 | 0.746 | 0.731 |
| | | Subject | 0.816 | 0.767 | 0.685 | 0.856 | 0.863 | 0.827 |
| | 16 weeks delayed progression | Mean | 0.663 | 0.955 | 0.951 | 0.767 | 0.716 | 0.926 |
| | | Subject | 0.876 | 0.993 | 0.992 | 0.945 | 0.913 | 0.978 |
| | 20% slowed progression | Mean | 0.663 | 0.715 | 0.593 | 0.757 | 0.716 | 0.804 |
| | | Subject | 0.860 | 0.831 | 0.777 | 0.895 | 0.892 | 0.924 |

# Effect estimates and coverage probability

**Table A3.** Effect estimates compared to ground truth using different methods across different simulated scenarios in Case Study 1. For the Time-PMRM, 'True effect quantified with method' was computed by minimizing the least-squares distance between the true active group trajectory and the placebo trajectory under the model. Scenarios where no ground truth is available due to model misspecification are marked with '—'.

| Effect | Method | True effect quantified with method | 300 patients per arm | | | 500 patients per arm | | |
|---|---|---|---|---|---|---|---|---|
| | | | Mean effect estimate | Median effect estimate | Standard deviation of effect estimates | Mean effect estimate | Median effect estimate | Standard deviation of effect estimates |
| No effect | cLDA (36 months) | 0 | 0.00 | -0.01 | 0.85 | 0.00 | 0.00 | 0.66 |
| | cLDA AUC difference | 0 | -0.18 | -0.41 | 15.72 | 0.13 | 0.45 | 12.07 |
| | PMRM proportional decline | 1 | 1.01 | 1.00 | 0.10 | 1.00 | 1.00 | 0.08 |
| | Time-PMRM (36 months) | 1 | 1.00 | 1.00 | 0.07 | 1.00 | 1.00 | 0.05 |
| | Time-PMRM proportional slowing | 1 | 1.01 | 1.00 | 0.06 | 1.00 | 1.00 | 0.05 |
| Stable symptomatic benefit | cLDA (36 months) | -0.78 | -0.83 | -0.83 | 0.82 | -0.79 | -0.79 | 0.65 |
| | cLDA AUC difference | -23.48 | -24.55 | -24.46 | 15.42 | -23.46 | -23.73 | 11.83 |
| | PMRM proportional decline | — | 0.97 | 0.97 | 0.10 | 0.97 | 0.97 | 0.08 |
| | Time-PMRM (36 months) | 0.94 | 0.95 | 0.94 | 0.07 | 0.95 | 0.95 | 0.05 |
| | Time-PMRM proportional slowing | — | 0.99 | 0.98 | 0.05 | 0.99 | 0.99 | 0.04 |
| Fading symptomatic benefit | cLDA (36 months) | -0.39 | -0.36 | -0.36 | 0.88 | -0.40 | -0.40 | 0.65 |
| | cLDA AUC difference | -18.78 | -18.74 | -18.74 | 15.95 | -18.65 | -18.70 | 12.28 |
| | PMRM proportional decline | — | 1.05 | 1.04 | 0.11 | 1.04 | 1.03 | 0.08 |
| | Time-PMRM (36 months) | 0.97 | 0.98 | 0.98 | 0.07 | 0.98 | 0.97 | 0.05 |
| | Time-PMRM proportional slowing | — | 1.01 | 1.00 | 0.06 | 1.00 | 1.00 | 0.04 |
| | cLDA (36 months) | -1.57 | -1.53 | -1.56 | 0.88 | -1.56 | -1.55 | 0.68 |
| 20% reduced decline | cLDA AUC difference | -23.19 | -22.86 | -23.40 | 16.19 | -23.29 | -23.06 | 12.32 |
| | PMRM proportional decline | 0.80 | 0.81 | 0.80 | 0.09 | 0.80 | 0.80 | 0.07 |
| | Time-PMRM (36 months) | 0.87 | 0.88 | 0.88 | 0.07 | 0.88 | 0.88 | 0.05 |
| | Time-PMRM proportional slowing | — | 0.91 | 0.92 | 0.06 | 0.91 | 0.92 | 0.04 |
| 4 months stable delay of progression | cLDA (36 months) | -1.20 | -1.25 | -1.24 | 0.87 | -1.22 | -1.20 | 0.65 |
| | cLDA AUC difference | -23.04 | -23.49 | -23.58 | 15.95 | -23.23 | -23.55 | 11.67 |
| | PMRM proportional decline | — | 0.85 | 0.84 | 0.10 | 0.85 | 0.85 | 0.08 |
| | Time-PMRM (36 months) | 0.90 | 0.90 | 0.90 | 0.07 | 0.90 | 0.90 | 0.05 |
| | Time-PMRM proportional slowing | — | 0.86 | 0.86 | 0.08 | 0.86 | 0.85 | 0.06 |
| 20% slowed progression | cLDA (36 months) | -2.16 | -2.18 | -2.17 | 0.86 | -2.17 | -2.14 | 0.66 |
| | cLDA AUC difference | -29.82 | -30.38 | -30.15 | 16.06 | -29.94 | -29.71 | 11.91 |
| | PMRM proportional decline | — | 0.73 | 0.73 | 0.09 | 0.73 | 0.72 | 0.07 |
| | Time-PMRM (36 months) | 0.82* | 0.82 | 0.83 | 0.07 | 0.82 | 0.82 | 0.05 |
| | Time-PMRM proportional slowing | 0.80* | 0.80 | 0.81 | 0.07 | 0.81 | 0.81 | 0.05 |
| Increasing slowing of progression | cLDA (36 months) | -2.16 | -2.21 | -2.24 | 0.88 | -2.15 | -2.15 | 0.64 |
| | cLDA AUC difference | -22.66 | -23.63 | -23.83 | 16.32 | -22.35 | -22.05 | 11.77 |
| | PMRM proportional decline | — | 0.75 | 0.75 | 0.10 | 0.76 | 0.75 | 0.07 |
| | Time-PMRM (36 months) | 0.82 | 0.82 | 0.82 | 0.07 | 0.82 | 0.82 | 0.05 |
| | Time-PMRM proportional slowing | — | 0.82 | 0.82 | 0.05 | 0.82 | 0.82 | 0.04 |

*Due to the slight difference between the linear interpolation used to generate the treatment effect and the spline interpolation used to estimate the treatment effect, the optimal least squares estimate is 0.82 while the actual slowing used to generate the data was 0.80

**Table A4** Coverage probabilities of nominal two-sided 95% confidence intervals for PMRM models in Case Study 1. For the Time-PMRM, 'True effect quantified with method' was computed by minimizing the least-squares distance between the true active group trajectory and the placebo trajectory under the model.

| Effect | Method | True effect quantified with method | Two-sided 95% confidence interval coverage probabilities | |
| --- | --- | --- | --- | --- |
| | | | 300 patients per arm | 500 patients per arm |
| No effect | PMRM proportional decline | 0 | 0.948 | 0.947 |
| | Time-PMRM (36 months) | 0 | 0.963 | 0.960 |
| | Time-PMRM proportional slowing | 0 | 0.897 | 0.919 |
| Stable symptomatic benefit | Time-PMRM (36 months) | 0.94 | 0.973 | 0.951 |
| Fading symptomatic benefit | Time-PMRM (36 months) | 0.93 | 0.964 | 0.955 |
| 20% reduced decline | PMRM proportional decline | 0.80 | 0.951 | 0.941 |
| | Time-PMRM (36 months) | 0.87 | 0.960 | 0.948 |
| 4 months stable delay of progression | Time-PMRM (36 months) | 0.90 | 0.961 | 0.960 |
| 20% slowed progression | Time-PMRM (36 months) | 0.82* | 0.962 | 0.959 |
| | Time-PMRM proportional slowing | 0.80* | 0.860 | 0.884 |
| Increasing slowing of progression | Time-PMRM (36 months) | 0.82 | 0.963 | 0.971 |

*Due to the slight difference between the linear interpolation used to generate the treatment effect and the spline interpolation used to estimate the treatment effect, the optimal least squares estimate is 0.82 while the actual slowing used to generate the data was 0.80

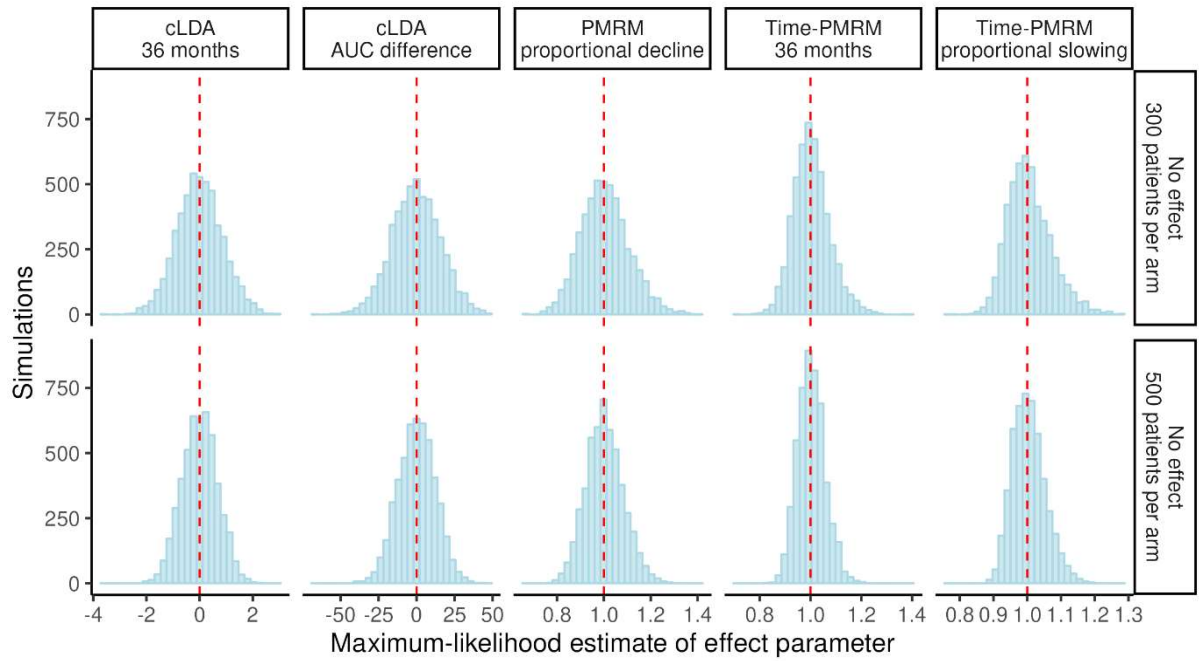

**Figure A1** Maximum-likelihood estimates of effect parameters of different methods across simulated null scenarios in Case Study 1.

# Testing proportional slowing of progression assumption

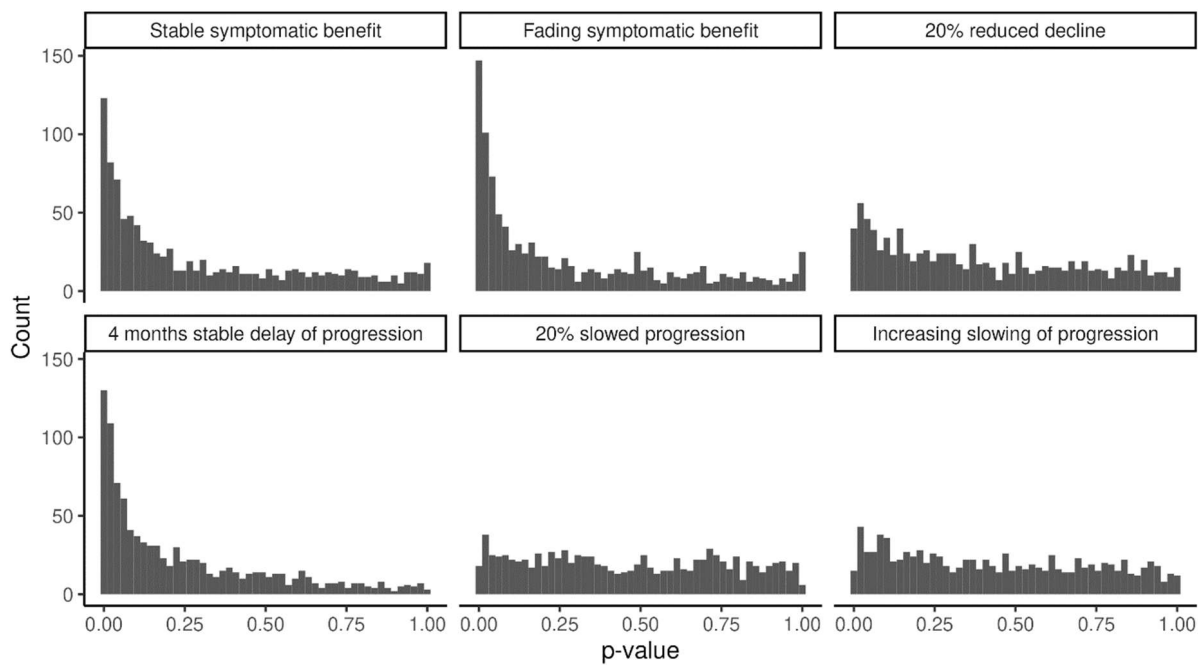

**Figure A2.** P-values for testing for violations of the proportional slowing of progression assumption across the 6 effect scenarios in case study 1 with 300 patients per arm. P-values are computed using a likelihood ratio test of the proportional slowing Time-PMRM against the Time-PMRM.

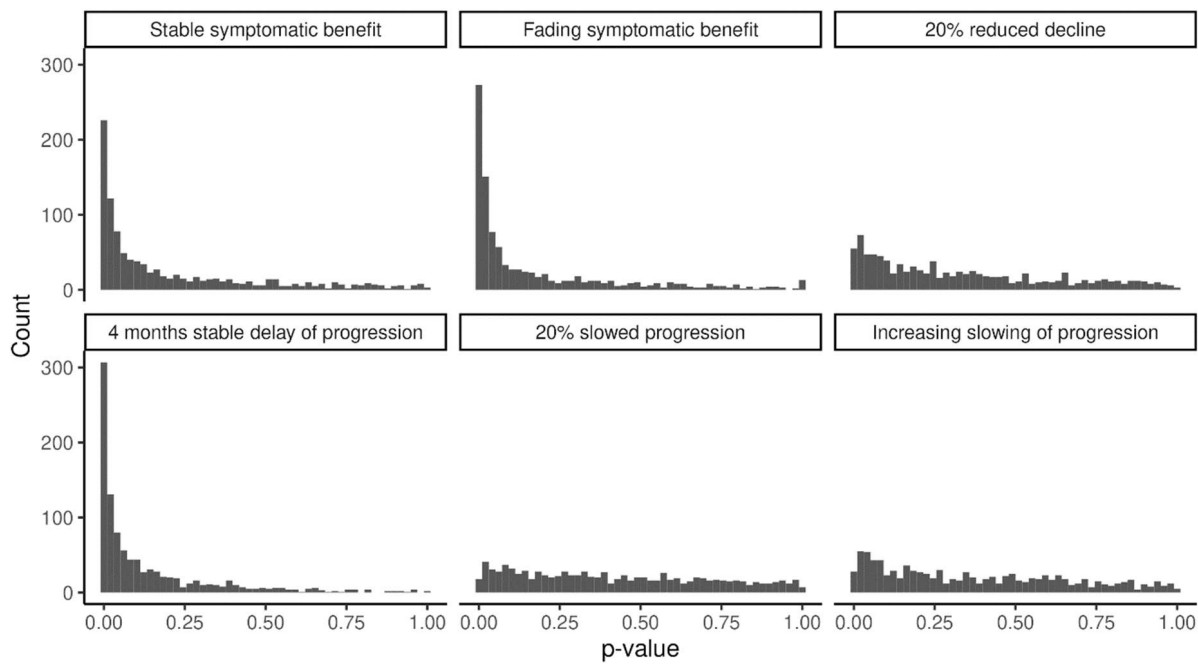

**Figure A3.** P-values for testing for violations of the proportional slowing of progression assumption across the 6 effect scenarios in case study 1 with 500 patients per arm. P-values are computed using a likelihood ratio test of the proportional slowing Time-PMRM against the Time-PMRM.